\documentclass[prb,article]{revtex4-1} 
\usepackage{amsmath}  
\usepackage{amsfonts} 
\usepackage{graphicx} 
\usepackage{color}

\begin{document}

\title{Incorporating Current Research into Formal Higher Education Settings using Astrobites}

\author{Nathan E. Sanders}
\email{astrobites@gmail.com}
\altaffiliation[permanent address: ]{Astrobites Collaboration, 1667 K Street NW, Suite 800
Washington, DC 20006, USA}
\affiliation{Astrobites Collaboration, 1667 K Street NW, Suite 800
Washington, DC 20006, USA}

\author{Susanna Kohler}
\affiliation{American Astronomical Society, Washington, DC 20006}

\author{Chris Faesi, Ashley Villar}
\affiliation{Harvard-Smithsonian Center for Astrophysics, Cambridge, MA 01238}

\author{Michael Zevin}
\affiliation{Northwestern University Department of Physics, Evanston, IL 60208-3112}

\author{and the Astrobites Collaboration}

\date{\today}

\begin{abstract}
A primary goal of many undergraduate- and graduate-level courses in the physical sciences is to prepare students to engage in scientific research, or to prepare students for careers that leverage skillsets similar to those used by research scientists.  Even for students who may not intend to pursue a career with these characteristics, exposure to the context of applications in modern research can be a valuable tool for teaching and learning.  However, a persistent barrier to student participation in research is familiarity with the technical language, format, and context that academic researchers use to communicate research methods and findings with each other: the literature of the field.  Astrobites, an online web resource authored by graduate students, has published brief and accessible summaries of more than 1,300 articles from the astrophysical literature since its founding in 2010.  This article presents three methods for introducing students at all levels within the formal higher education setting to approaches and results from modern research.  For each method, we provide a sample lesson plan that integrates content and principles from Astrobites, including step-by-step instructions for instructors, suggestions for adapting the lesson to different class levels across the undergraduate and graduate spectrum, sample student handouts, and a grading rubric.
\end{abstract}

\maketitle

\section{Introduction}

Scientific literature simultaneously serves two roles: it is the primary means by which we communicate our results to our colleagues, and it provides a historical record of progress in our field. But the literature should also play a critical third role: the introduction to our profession for the next generation of researchers.  While textbooks also play an important role in this introduction, by their nature they do not encompass the most recent research results.  Moreover, textbooks cannot expose the process and motivations of science in the same way the primary source material of the scientific literature does.

Unfortunately, success in the first two roles often inhibits success in the third.   Effective communication between peers relies on shorthand, established jargon.  Experts are able to unspool references by using their familiarity with the context, and they can fluidly interpret discipline-specific vernacular and units. To establish a legacy continuous across generations of scientists, the literature must make passing reference to decades of research results, presenting an impenetrable web of dense manuscripts to newcomers.  This common vocabulary of ideas--thoughts, terms, and historical findings--is integral to concise communication among specialists but obscures meaning and clarity to students of the field.  

Many students' first contact with the literature presents a barrier rather than a facilitator to a career in research.   The process starts when an undergraduate approaches a professor or other researcher, asking to take part in a research project for the first time.  For a mentor eager to get a first-time researcher up to speed, it can seem convenient and straightforward to provide the student with a set of the canonical, comprehensive written works in the field.  But without prior exposure to the literature, the first paper sets a rather high activation energy for students to surpass to become an active contributor to the field.  

\subsection{Integrating Research and Education}

Undergraduate research experiences have been shown to be an integral step in the STEM career path\cite{seymour04,willison07}.  Lowering barriers to participation in research facilitates the pursuit of several goals. It may make the practice of scientific research more accessible to students from backgrounds traditionally under-represented in the sciences and lead to higher retainment of such students.\cite{ceci09,hernandez13}.  Furthermore, it could improve science self-efficacy\cite{robnett15} among emerging researchers and the productivity of early-career scientists. 

While facility with the research literature in the field is unambiguously critical to the success of academic researchers in any scientific field, the comprehension, communication, and analytical skills associated with that facility are broadly applicable across nearly all occupations pursued by trained physicists.  When the Joint Task Force on Undergraduate Physics Programs of the American Physical Society (APS) and American Association of Physics Teachers (AAPT) released their final report in 2016,\cite{jtupp} three of the four consensus skills they identified as instrumental to preparing physics students for careers in research or the modern workforce were related to this topic: ``understanding how science and technology are used in real-world settings,'' ``writing and speaking well'', and ``understanding the context in which work is now done.''

In particular, there is an increasing recognition of the importance of such``non-cognitive'' or ``success-critical'' skills in graduate admissions.  While traditional measures of student success like the Graduate Record Exam and grade point average have long been used as predictors of future performance in research, there is growing evidence that success-critical skills such as student adaptability and motivation are better correlated with long term success, particularly among underrepresented groups.\cite{kent}  Similarly, several authors have reported relationships between gender and contextual learning,\cite{murphy06} the presentation of canonical knowledge and fundamental concepts in the context of real world applications such as modern research projects.

Instructors must use a variety of techniques to introduce students to the literature in their field and enable them to benefit from the knowledge recorded there in the way that experienced researchers are able to do.  These techniques include strategies for database searching,\cite{miller09} online computational simulations to engage students with modern competitive models for not-yet-settled subjects of research,\cite{moldenhauer13,popp15} and student-driven research and publishing simulation.\cite{eagles16}  By integrating these techniques into the formal, classroom education setting, all students can be instilled with these skills at the same time they are building their content knowledge in the field. Here we focus on the suitability of Astrobites, a free web resource that provides scaffolds for students seeking to read journal papers in astrophysics for the first time, for use in formal education courses to introduce students to the research literature.

\subsection{Astrobites}

\begin{figure}[h!]
\centering
\includegraphics[width=6in]{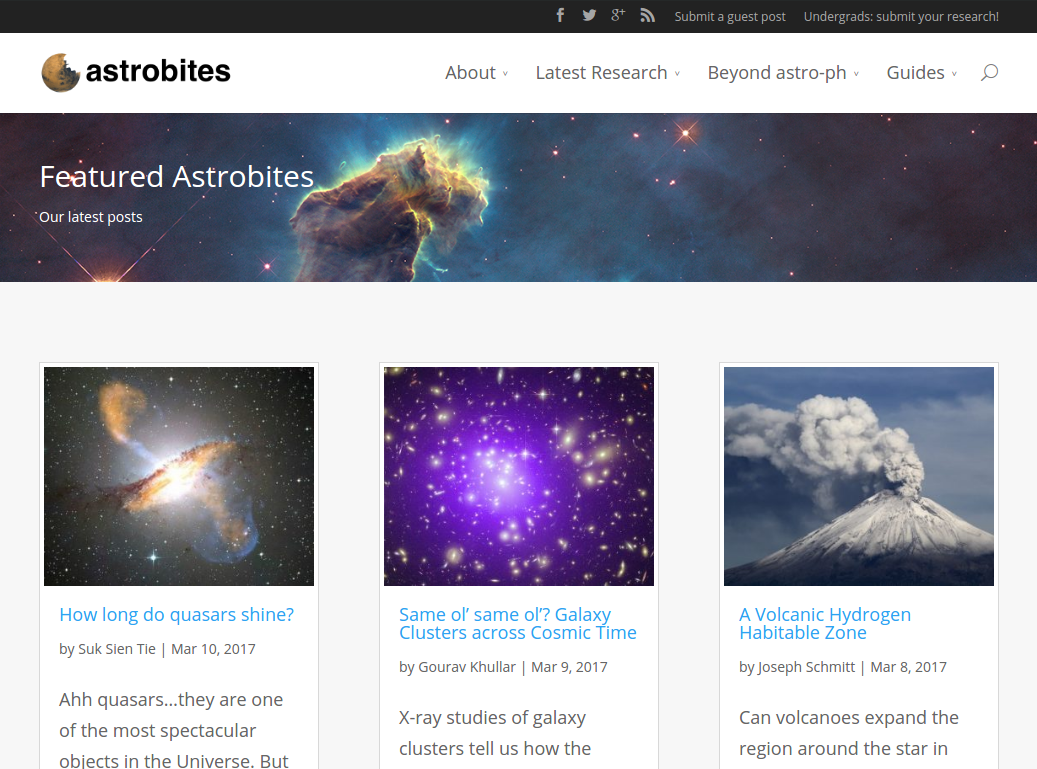}
\caption{Screenshot of the front page of the Astrobites website as it appeared on March 10, 2017.}
\label{FigAstrobites}
\end{figure}

Astrobites (\url{<http://astrobites.org/>}; Figure~\ref{FigAstrobites}) is the reader's digest of the astrophysical literature.  The Astrobites website publishes brief and accessible summaries of recent research papers targeted at an undergraduate-level audience.  In addition to summarizing longer technical reports, translating unfamiliar jargon, and providing links to additional information on important concepts, Astrobites provides crucial context to readers.  While articles in peer reviewed journals typically assume knowledge of decades of literature, often referring to an entire sub-field of literature or a complex methodological specification by the citation of a single paper, Astrobites articles provide brief discussions of the essential past work and current motivations for research to help introduce young scientists to this context.  Astrobites does assume a working knowledge of the basic principles of mechanics, electromagnetism, optics, and other relevant undergraduate-level subjects in physics, alleviating the need to repeatedly rehash fundamental topics.  Astrobites has published more than $1,300$ research paper summaries as well as hundreds of other articles on career navigation, interviews with researchers, technical tutorials on popular software packages, and more.  Through a partnership with the American Astronomical Society (AAS), Astrobites also supplies intensive student-driven coverage of the bi-annual AAS meetings.\cite{astrobitesAAS229}  All content on Astrobites is free to access and paper summaries link back to the original technical publication.

Since its founding in 2010, Astrobites has pioneered a unique collaborative author model driven by the volunteer efforts of graduate students around the world.  Each day, one graduate student member of the Astrobites collaboration drafts an original summary of a recent research publication that has appeared in a peer reviewed journal or on the astrophysical preprint server \textit{astro-ph}\cite{astroph}, while another member of the collaboration is assigned to edit this work.  The Astrobites editorial model has proven highly sustainable and extensible to other scientific fields.  Astrobites has sister organizations publishing in Spanish (\url{<https://astrobitos.org>}) and on the subjects of particle physics (\url{<https://particlebites.com>}), oceanography (\url{<http://oceanbites.org>}), and more.  Many sister sites of Astrobites were founded by graduate student attendees of the Communicating Science workshop, ComSciCon (\url{<http://comscicon.com>}), an annual national conference on leadership in science communication for graduate students organized in part by administrators of the Astrobites collaboration.

\subsection{Lesson Plans}

In this work, we present three original lesson plans that use Astrobites as a resource for instructors to integrate modern research methods and results into their teaching at the undergraduate- or graduate-level.  As part of each lesson plan, or ``method,'' we provide specified learning objectives, step-by-step instructions including optional or extra credit assignments, strategies for adapting the lessons to different course levels, and supporting materials.  The lesson types and the teaching strategies we recommend frequently prompt online student collaboration and discussion of conceptual questions related to course material, behaviors that have been shown to correlate with student success in physics higher education settings.\cite{kortermeyer06}  The first (Method~1) makes use of periodic reading assignments to promote student familiarity with the astrophysical research, and uses structured online questions to encourage and check student understanding.  Method~2 prompts students to do original research on an astrophysical topic using Astrobites and other resources for delivery in a written paper or oral presentation.  Optional assignments have students construct conceptual graphs or historical timelines of relevant research to visually communicate what they've learned from their scaffolded review of the literature.  The final lesson type (Method~3) asks students to write their own research paper summary within the parameters and style of an Astrobite, optionally including interactivity through construction of a collaborative bibliography and peer feedback.

While we have previously written about the importance of integrating research into the undergraduate classroom and the opportunity for educators to make use of Astrobites as a resource,\cite{sanders12} this is the first time that lesson plans have been developed and made available for educators to immediately use in their classes.  In this work, we further supplement these strategies by specifying learning objectives, instructions, and materials for each lesson type.  These materials were first workshopped with 25 students and educators at the 229th AAS meeting in Grapevine, TX.  Many of the specific teaching strategies we present here were suggested by these participants and developed during and after that session.  

\section{Method 1: Periodic Reading Assignments}

Students are asked to read an assigned Astrobites article and respond to guided questions that test reading comprehension and conceptual understanding.  Students are graded on their responses to the questions, which can be gathered electronically through an online form.  The questions are discussed in class to promote greater understanding.  

This assignment is appropriate for any course level, including introductory level classes, and is reproducible.  It could be done, for example, once per week or once per class period.  These readings can be used as pre-lab assignments to connect lab activities to current research.\cite{maryland}

\subsection{Learning Objectives}

\begin{itemize}
	\item \textbf{Reading comprehension:} Students will gain the ability to extract information from new reading, including identifying connections to course curriculum or previous readings.

	\item \textbf{Conceptual understanding:} Students will strengthen their understanding of course curriculum by integrating information from new readings. They will demonstrate learnings from course content or previous readings by answering guided questions.

	\item \textbf{Literature familiarity:} Over several such assignments, students will develop familiarity with the astronomical literature.  They will form an understanding of active disciplines of research and the ability to draw connections to the course curriculum or their own research interests. 
\end{itemize}

\subsection{Instructions}

\begin{enumerate}

\item \textbf{Instructor selects an Astrobites article and formulates 2-4 guided questions.}

Guided questions are intended to encourage mindful student reading of the assigned article and check student understanding of major concepts.  We suggest questions that can be answered by straightforward application of concepts learned in previous lessons to newly-encountered subject matter from the reading, for example applying an understanding of the Doppler effect to describe the observational effect of the orbit of Alpha Centauri on its radial velocity.  Questions typically either would not include calculations, or they would focus only on symbolic manipulation or scaling relations.

\item \textbf{Instructor distributes Astrobites article to students along with guided questions in the form of a fillable Google Form.}

Students submit responses to guided questions.

In Appendix~\ref{AppendixM1}, we provide a sample Google Form that can be used to gather student responses to questions electronically.

\item \textbf{(Optional or extra credit) Students leave a comment on the assigned Astrobite with a question or thought about the article.}

Astrobites authors track article comments, so students will usually receive a timely response to their comment.  All Astrobites authors are practicing scientists (graduate students or recent grads) and so can provide additional subject matter insight or answer questions about the practice of science in their reply.

\item \textbf{(Optional or extra credit) Students write a brief essay contrasting the original journal paper to an institutional press release about the same paper.}

The instructor selects an Astrobites article about a paper that had an associated press release.  Students are asked to compare and contrast the scientific findings emphasized in the paper versus the press release, identify the scientific concepts that were explained for non-technical audiences in the press release, and comment on how the uncertainty and significance of the work were presented in the press release.  Instructors can use the Astrobite as a resource to provide additional context and perspective for this assignment.

\item \textbf{Instructor evaluates responses and scores according to rubric below.}

\item \textbf{Instructor reviews guided questions during next class period.  }

One or more students are asked to provide their answers to each question.  Instructor guides student discussion towards accurate understanding, as appropriate.

\end{enumerate}

\subsection{Adaptation to Different Course Levels}

\begin{itemize}
\item \textbf{Entry level undergraduate} \newline Instructors should:
	
	\begin{itemize}
	\item Carefully select articles that are closely related to the current subject matter at hand and that do not introduce new concepts that have not yet been covered in the course.
	\item Assign readings less frequently and with several days before due date to promote careful reading.
	\item Preferentially pose questions that test understanding of course content more so than full reading comprehension.
	\item Focus on questions with straightforward interpretations and use open-ended questions less frequently.
	\item Challenge students to interpret quantitative statements from the paper by generating questions that require students to make real-life connections. As an example, ``How many times larger in radius is your university campus compared to a neutron star?''	
	\item Prompt students to reflect on what they've read by collecting brief responses to the open-ended prompt, ``I used to think...  Now I think...''
	\item Ask students to contribute a certain amount of time to a citizen science project\cite{zooniverse} with connections to the paper they have reviewed.
	\end{itemize}

\item \textbf{Upper level undergraduate} \newline Instructors should:

	\begin{itemize}
	\item Give assignments more frequently to encourage students to develop strong familiarity with the literature.
	\item Try collecting feedback from students with questions like, ``What sentence from the article did you not have enough information to understand?’’ or ``What subject from the article would you like to learn more about in class?’’  Student responses can be reviewed before the next lesson and used to inform instruction.
	\item Reinforce conceptual understanding of physical formulae from the course by using questions that ask students to develop order-of-magnitude arguments that demonstrate consistency with calculations from the paper.
	\item Ask students to ``update'' a relevant section of their textbook by adding information from the current research article, using a short written response to the questionnaire or a brief in-class discussion.
	\item Use the questionnaire to ask students to annotate a figure from the paper or a related diagram with a caption and/or axes labels using their own words to demonstrate comprehension. 
	\end{itemize}

\item \textbf{Graduate level} \newline Instructors should:

	\begin{itemize}
	\item Preferentially assign recent articles covering new results to challenge students to draw connections to current topics without being pulled off course by less relevant content.
	\item Preferentially pose questions that invite students to extend beyond the subject matter of the article at hand or apply their understanding to personal research experience.
	\item Build student understanding of physical formulae from the course curriculum by posing questions that call on students to reproduce calculations from the paper.
	\item Use open-ended question formats to explore student perspectives.
	\item Occasionally ask students to go beyond the Astrobites summary by asking questions that direct them towards the source paper.
	\item Use this assignment as a launching point to a Method~2 or 3 assignment in the same topic area.
	\end{itemize}

\end{itemize}

\subsection{Materials}

\begin{figure}[h!]
\centering
\includegraphics[width=6in]{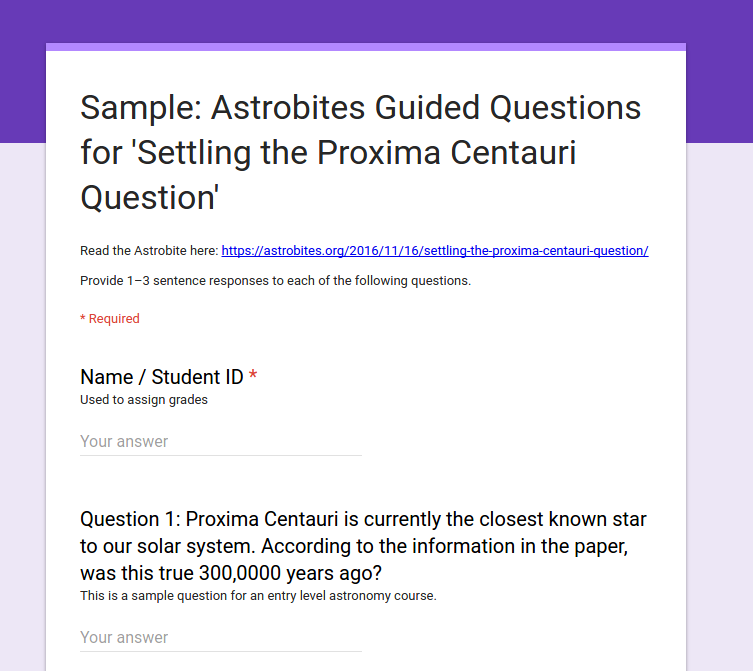}
\caption{Screenshot of the questionnaire for collecting responses to guided questions associated with Method~1 and provided in Appendix~\ref{AppendixM1}.}
\label{FigM1questionnaire}
\end{figure}

Appendix~\ref{AppendixM1} provides a sample grading rubric and an online form template for collecting responses to student guided questions for instructors to modify to their own course setting.  Figure~\ref{FigM1questionnaire} displays a screenshot of the first part of the sample questionnaire for guidance as to its form and content.

\section{Method 2: Student Research Project}

Students are asked to select a research topic and then identify and read several Astrobites articles related to that topic.  Students then prepare a written paper or class presentation based on this independent research.  In advanced courses, students can be asked to read the source material (original paper), so that the Astrobites article serves as scaffolding to introduce them to that material.  A set of optional research tasks asks students to construct an annotated bibliography of their reading and/or a concept graph that links topics from the course curriculum to the modern research.  While we suggest modifications to make use of this method at introductory undergraduate- through graduate-levels, this method is perhaps best suited to the upper-level undergraduate course.

\subsection{Learning Objectives}

\begin{itemize}
\item \textbf{Reading comprehension:} Students will gain the ability to extract information from new reading, including identifying connections to course curriculum or previous readings.

\item \textbf{Synthesis:} Students will gain the ability to synthesize concepts and information from a variety of sources into an original work in a presentation or paper format.

\item \textbf{Interpreting data:} Students will gain the ability to extract information from data visualizations and statistical graphics.  Students integrate figures from published research into their projects and use them to support their descriptions or arguments.
\end{itemize}

\subsection{Instructions}

\begin{enumerate}

\item \textbf{Students select a research topic.}

The students can select their topic from an instructor-provided list of topics specific to the course curriculum, from the most popular tagged topics within our daily paper summaries category,\cite{mostpopular} or from elsewhere.

\item \textbf{Students identify 2-4 Astrobites articles related to the topic.}

(Optional) Utilize the student research to construct an annotated bibliography for the course.  Instructors can use a Google Form like the example in Appendix~\ref{AppendixM2} to collect a response from each student to document their research on each Astrobite.  After the initial research assignment (Step 2), instructors can then make the filled responses available to all students as a resource to help with the remainder of the project (Steps 3+).  Example Google forms and filled responses are provided in Appendix~\ref{AppendixM2}.

\item \textbf{(Optional or extra credit) Ask students to construct a concept graph.}

Students write down a list of significant concepts from the curriculum that are prerequisite to understanding the present topic.  The list of concepts could be culled, for example, from the table of concepts of their course text.  The student then constructs a dependency graph showing how the topics link together. Examples of similar concept graphs are provided by HyperPhysics.\cite{hyperphysics}

\item \textbf{(Optional or extra credit) Ask students to construct a research timeline.}

Students consult the introductions of the source papers associated with their Astrobites selections.  They cross-reference the citations in the sources to reconstruct a sequence of major milestones in research related to their topic.  

The Astrophysical Data System (ADS) Bumblebee\cite{bumblebee} Author Network tool can also help students.  Students can search for a subject, sort in descending order by citation, and then view the author network to get a view of major collaborations and how their contributions impacted the field over time.

\item \textbf{Students complete their project assignment.}

	\begin{itemize}
	\item \textbf{Student presentation path:} Students are asked to prepare a brief ($5-10$ minute) group or individual presentation about the topic they selected, to deliver in class.

	Instructors encourage students to build skills in interpreting data visualizations by asking them to include and explain one or more figures from the Astrobites source papers in their presentation.

	\item \textbf{Student paper path:} Students are asked to prepare a short ($4-10$ pages) written paper related to the topic they selected.  

	The instructor should give guidance as to how the paper should compare to a typical Astrobite or to the published source papers in terms of accessibility (how easy it should be for, for example, someone who has not taken your course to understand) and level of detail.  A diagram like Figure~\ref{FigExpectations} can be helpful.

	\end{itemize}

\begin{figure}[h!]
\centering
\includegraphics{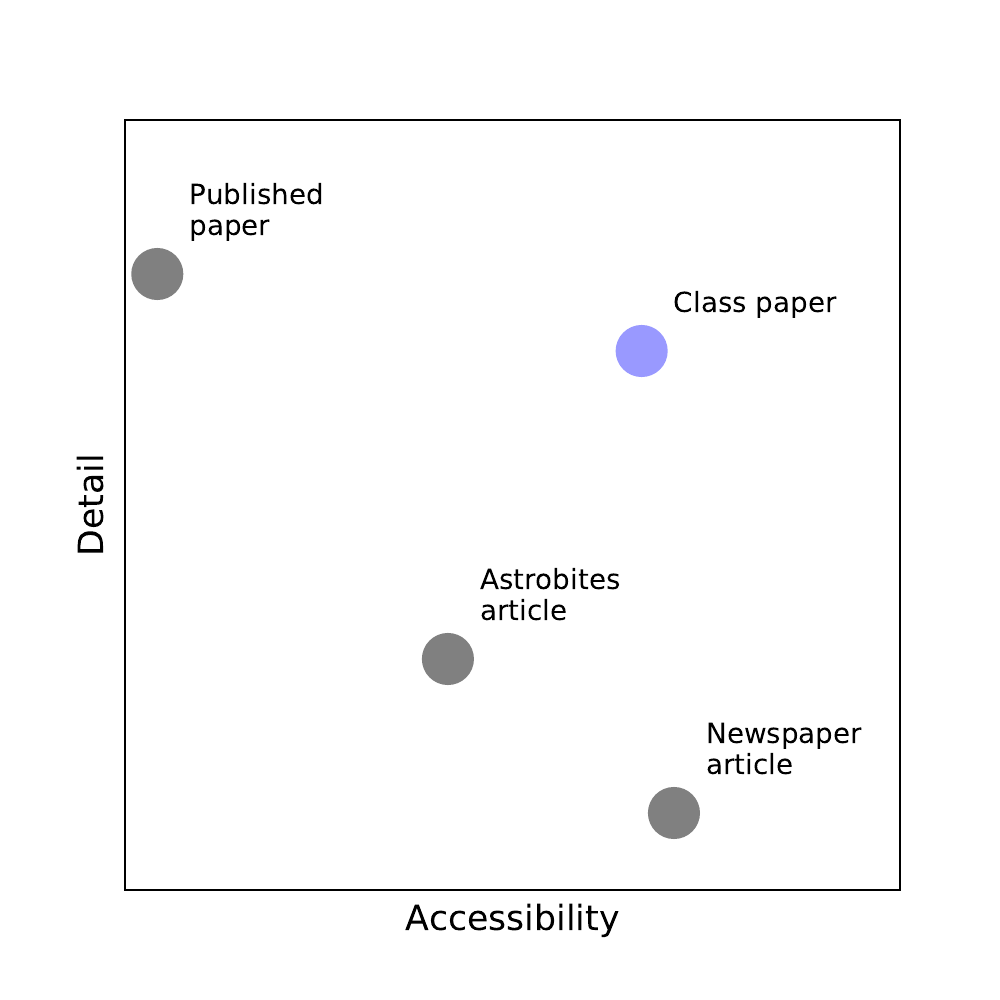}
\caption{Sample diagram to help communicate accessibility and detail expectations to students for the student paper assignment associated with Method~2.}
\label{FigExpectations}
\end{figure}

\item \textbf{Instructor reviews responses and scores according to rubric in Appendix~\ref{AppendixM2}.}

\end{enumerate}

\subsection{Adaptation to Different Course Levels}

\begin{itemize}
\item \textbf{Entry level undergraduate} \newline Instructors should:

	\begin{itemize}
	\item Since Astrobites articles generally link to several past articles to establish foundational concepts, encourage students to make use of those links if they have trouble constructing a bibliography through their own searches.
	\item For students that may major in or have experience in other disciplines, ask them to compare the process of astronomical research to those other fields.  What aspects of the research process in astronomy are familiar, or seem surprising?  
	\end{itemize}

\item \textbf{Upper level undergraduate}

	\begin{itemize}
	\item Challenge students to dive into the original source articles for each Astrobite.  Instructors can use the ``annotated bibliography'' form to ask students to provide additional information extracted from the source article.
	\end{itemize}

\item \textbf{Graduate level} \newline Instructors should:

	\begin{itemize}
	\item Use the ``annotated bibliography'' form to invite students to augment the reading from Astrobites with their own knowledge.  Instructors can append a question to the form that prompts students to provide additional context that they feel would add understanding to what was presented in the Astrobite.
	\item Ask students to criticize the body of research they've investigated.  Are there lines of evidence that are more or less rigorous?  Are there alternative theories that deserve more or less credence?  What are the caveats to the observational, statistical, or theoretical approaches used in the field?
	\end{itemize}

\end{itemize}

\subsection{Materials}

Appendix~\ref{AppendixM2} provides a sample grading rubric and student handout for generating a collaborative bibliography that instructors can modify to their own course setting.

\section{Method 3: Student Writing Assignment}

In this project, designed with upper-level undergraduate or graduate classes in mind, students write their own Astrobites-like article to synthesize and summarize the content of one or more research papers.  As the outcome is a set of brief written summaries based on substantial amounts of student research, this assignment helps to build research, literature understanding, and communication skills without subjecting instructors to a high burden in reading long pieces of student work.  A sample handout provides guidance as to the form, content, and style of the written pieces.

\subsection{Learning Objectives}

\begin{itemize}
\item \textbf{Reading comprehension:} Students will gain the ability to extract information from new reading, including identifying connections to course curriculum or previous readings.
\item \textbf{Composition:} Students will gain the ability to synthesize concepts and information from a variety of sources into an original written work.
\item \textbf{Communication:} Students will gain the ability to convey ideas and knowledge to an audience that may be different in age, training, perspective, or experience than the student.
\item \textbf{Interpreting data:} Students will gain the ability to extract information from data visualizations and statistical graphics.
\end{itemize}

\subsection{Instructions}

\begin{enumerate}

\item \textbf{Introductory Reading Assignment.}

Students read an introductory source article and associated Astrobite, following e.g. the approach of Method~1.  Instructors should select an article with a topic and summarization that they deem appropriate for the course.  This serves to introduce students to the summarization style and the level of outside context they should add as authors.

\item \textbf{Students select a source article.}

The instructor can provide a list of research papers tailored to the course curriculum, or students can be asked to identify their own.  Especially for lower-level courses, articles that have previously been the subject of Astrobites posts can be a good fit (see Section~\ref{AdaptM2}).

Students should be given constraints on the source material, such as articles appearing in a certain journal, or on the preprint server the arXiv.  Students should submit their selection to the instructor to verify that it is appropriate before they begin drafting their piece.

\item \textbf{Students write a first draft of their article.}

The project description handout in Appendix~\ref{AppendixM3} contains an example writing prompt and example article template.  

The most important choice for the instructor is what audience to direct students to write for.  A clear focus in audience should set strong expectations for the content level of the piece.  Astrobites articles are written for an undergraduate-level audience of physics or astronomy majors.  Instructors might ask students to target an audience of their peers, or perhaps at one or two levels below their current status.

As in this example, we recommend setting a fairly narrow word count range and directing students to adhere to it, as this project is meant to help students build experience writing in a concise and clear format.  We also recommend asking students to include $1-2$ figures from the source paper in their article with captions written in their own words. This will strengthen their ability to understand and convey the significance of graphics and data visualizations in scientific writing.

\item \textbf{Students exchange drafts for peer editing.}

Just like all Astrobites articles are proofread for content and style by another graduate student in our collaboration, we recommend that students provide peer feedback on their classmates' writing.  Instructors can supply the peer editing rubric provided Appendix~\ref{AppendixM3} to guide student feedback.

If possible, instructors are encouraged to hold peer review sessions in class, so instructors can provide oversight and help ensure uniformly thoughtful and constructive feedback.  The provided feedback rubric asks peers to force rank areas for improvement and answer guiding questions to encourage constructive critical feedback.

\item \textbf{Students submit final drafts of their article.}

Instructors review articles and scores according to the rubric in Appendix~\ref{AppendixM3}.

\item \textbf{(Optional) Submit the article as a guest post to Astrobites.}

If the target audience for the piece is set by the instructor at or near Astrobites' undergraduate-level target, then interested students are welcome to submit their articles to Astrobites for possible publication as a guest post.  Students are asked to follow the instructions on our website (\url{<https://goo.gl/54zKGs>}) to submit and clearly label the submission as one associated with a class, mentioning the university, instructor, and course.

Submissions on recent articles are preferred.  Submissions in html or a word processing format (Google Doc, MS Word, OpenDocument) are preferred.

\end{enumerate}

\subsection{Adaptation to Different Course Levels}\label{AdaptM2}

\begin{itemize}
\item \textbf{Entry level undergraduate} \newline Instructors should:
	\begin{itemize}
	\item Consider having students instead read and summarize in their own words articles that already have previously-written summaries.  Students at this level will not likely be able to read and summarize articles directly from the scientific literature on their own.  Valuable resources include past Astrobites subject articles or the historical articles from Marcia Bartusiak's \textit{Archives of the Universe}.\cite{bartusiak}
	\item Add some structure to the assignment by asking students to focus on a very specific aspect of the article that ties into the course curriculum, such as a particular physical law or discovery.
	\end{itemize}
\item \textbf{Upper level undergraduate}
	\begin{itemize}
	\item Consider asking students to submit their articles in a particular digital format of your choice such as a latexed PDF or html to build technology skills relevant to communication and publication.
	\end{itemize}
\item \textbf{Graduate level} \newline Instructors should:
	\begin{itemize}
	\item For a sense of the commitment involved, consider that the time Astrobites' own graduate students typically spend reading and summarizing an article can vary from $\sim3-8$ hrs.
	\item Ask students to incorporate salient, related research such as the presentation of a recent colloquium speaker at their institution or the students' or advisors' own work.  Optionally or for extra credit, the student can be invited to interview the researcher to add additional context to their writing.	
	\end{itemize}
\end{itemize}

\subsection{Materials}

Appendix~\ref{AppendixM3} provides a sample grading rubric and student handouts for the project description, article template, and peer editing rubric, for instructors to modify to their own course setting.  

Instructors may also wish to point students towards Astrobites' guide to reading scientific articles in astronomy\cite{gifford11} to support them as they compile research for their papers.

In addition, below we provide an assortment of sample articles from previously published Astrobites for use as classroom examples:

\begin{itemize}
\item \textbf{Solar system formation:} \url{<https://goo.gl/tjI8ZY>}
\item \textbf{Stars:} \url{<https://goo.gl/5RFZsv>} 
\item \textbf{Dark matter:} \url{<https://goo.gl/NltF4j>} 
\item \textbf{Compact objects:} \url{<https://goo.gl/kCUV1w>} 
\item \textbf{Black holes:} \url{<https://goo.gl/2xGWej>} 
\item \textbf{Exoplanets:} \url{<https://goo.gl/GeVuRb>} 
\item \textbf{High-z galaxies:} \url{<https://goo.gl/jGXmZP>} 
\end{itemize}

\section{Conclusion}

We have presented three original methods for introducing young scientists at the undergraduate- and graduate-level to the research literature, each using the online web resource Astrobites.  We have argued that familiarity with the research literature is an instrumental part of the development of young scientists, whether they are pursuing careers in academic research or in careers that leverage skills similar to those used by practicing researchers.  As a freely accessible, daily digest of articles selected from the research literature, Astrobites is a natural tool for instructors to use in introducing their students to this critical body of knowledge.

We have presented lesson plans, including learning objectives, optional extra credit assignments, recommended adaptions to different course levels, and supporting materials for three instructional methods: periodic reading assignments, an original research project, and a research synthesis assignment.  In the future, we hope to work with instructors to study the impacts of using Astrobites in their classrooms.  While a longitudinal study of career outcomes based on exposure to the methods presented here would take decades to perform and be very difficult to control, a controlled study of facility with primary sources (can students comprehend and critically analyze papers selected from the modern literature?), perceptions (how comfortable do students feel consulting the literature for information?), and/or literature reading habits (how many original research papers do students read per month?) would be more straightforward.  Ideally, we would perform this research across a range of student levels, from introductory undergraduate- through graduate-level.  As we believe the existing level of access to experiences in research is greater at research-focused universities, we would like to study this impact in a range of institutional settings.

\appendix

\section{Materials for Method 1}\label{AppendixM1}

\subsection{Grading Rubric}

Table~\ref{TabM1GR} provides a sample grading rubric for the student reading assignments associated with Method~1.  The scores are meant to be assigned per question.
	
\begin{table}[h!]
\centering
\caption{Grading rubric for Method 1}
\begin{ruledtabular}
\begin{tabular}{p{0.1\linewidth} p{0.35\linewidth} | p{0.55\linewidth}}
Score & Completeness & Content \\
\hline	
0 & Questions not answered or partially answered & Student does not address all questions or demonstrates partial or no understanding of underlying concepts. \\
1 & Question completely or mostly answered & Student demonstrates some understanding of concepts underlying guided questions, but may have major factual inaccuracies or logical inconsistency. \\
2 & Question completely answered, with $1-3$ sentences & Student demonstrates strong overall understanding of concepts underlying guided questions, though may yet have some factual inaccuracies or logical inconsistency. \\
\label{TabM1GR}
\end{tabular}
\end{ruledtabular}
\label{bosons}
\end{table}

\subsection{Sample Student Handout}

We supply a sample student handout with example questions for distribution to students via a Google Form, which students can also use to submit their responses to the instructor.  Responses received will be formatted in an editable Google Spreadsheet for the instructor with timestamps for each submission.  The sample handout is available here: \url{<https://goo.gl/d3IAy9>}

\section{Materials for Method 2}\label{AppendixM2}

\subsection{Grading Rubric}

Table~\ref{TabM2GR} provides a sample grading rubric for the student reading assignments associated with Method~2.  The rubric corresponds to a total of 30 points.
	
\begin{table}[h!]
\centering
\caption{Grading rubric for Method 2.}
\begin{ruledtabular}
\begin{tabular}{p{0.1\linewidth} p{0.3\linewidth} | p{0.3\linewidth} | p{0.3\linewidth}}
Score & Scope & Accuracy & Communication \\
\hline	
$0 - 3$ & Total number of sources consulted or project length significantly less than assigned. & Significant factual or conceptual errors presented.  An expert in the selected topic would not have agreed with fundamental points made. & Paper / presentation incomplete, not comprehensible, and/or not aligned to course standards.  A non-expert would not have learned from it. \\
$4 - 7$ & Student integrated fewer sources than expected or did not cover as many aspects of the selected topic as expected. & An expert in the topic may have pointed out a few, minor factual or conceptual errors presented.  Student presentation would not have significantly enhanced their peers' understanding of the selected topic. & Paper / presentation was largely informative for peers in the course, but had some flaws in explanation or depth that inhibited understanding.  A non-expert would have learned about some elements of the selected topic. \\
$8 - 10$ & Student consulted and integrated at least the expected number of Astrobites articles and other sources.  Student comprehensively discussed aspects of the selected topic at the level expected for the course. & No factual or conceptual errors presented.  Student presentation aided their peers' understanding of the selected topic. & Paper / presentation was strongly coherent, informative, and clearly understood by peers in the course.  A non-expert would have learned substantially from it. \\
\label{TabM2GR}
\end{tabular}
\end{ruledtabular}
\label{bosons}
\end{table}

\subsection{Sample Student Handout}

We supply a sample student handout for generating a collaborative bibliography via student submissions to a Google Form.  Student responses received will be formatted in an editable Google Spreadsheet for the instructor.  The sample handout is available here: \url{<https://goo.gl/kYTQbT>}.  An example of filled responses associated with the form is available here: \url{<https://goo.gl/oTj9cc>}.

\section{Materials for Method 3}\label{AppendixM3}

\subsection{Grading Rubric}

Table~\ref{TabM3GR} provides a sample grading rubric for the student reading assignments associated with Method~3.  The rubric corresponds to a total of 30 points.
	
\begin{table}[h!]
\centering
\caption{Grading rubric for Method 3.}
\begin{ruledtabular}
\begin{tabular}{p{0.1\linewidth} p{0.3\linewidth} | p{0.3\linewidth} | p{0.3\linewidth}}
Score & Scope & Content & Communication \\
\hline
$0-3$ & Article incomplete or significantly different in length than the requirement (shorter or longer).  Does not cite other articles or resources beyond the subject. & Article has significant factual inaccuracies. & Article is difficult to understand due to typographical or grammatical issues, or because it is written at a level incongruous with the target audience. \\
$4-7$ & Article deviates somewhat from length requirement (shorter or longer).  Occasionally cites other articles or resources beyond the subject. & Article may have minor factual inaccuracies.  Article provides little additional context beyond what was explicitly mentioned in the subject. & The article would be understandable by a member of the target audience, but its value would be impaired by moderate typographical, grammatical, or content level inconsistencies. \\
$8-10$ & Article in line with word count requirement.  Frequently cites other articles or resources beyond the subject. & Article has no factual inaccuracies and provides significant additional context beyond that explicitly mentioned in the subject. & Article is well edited and written clearly.  It would be readily understandable by a member of the target audience. \\
\label{TabM3GR}
\end{tabular}
\end{ruledtabular}
\label{bosons}
\end{table}

\subsection{Sample Student Handout}

We supply student handouts for several different purposes.  These are supplied as editable Google Documents that educators can copy and modify for their own classroom use.

\begin{figure}[h!]
\centering
\includegraphics[width=4in]{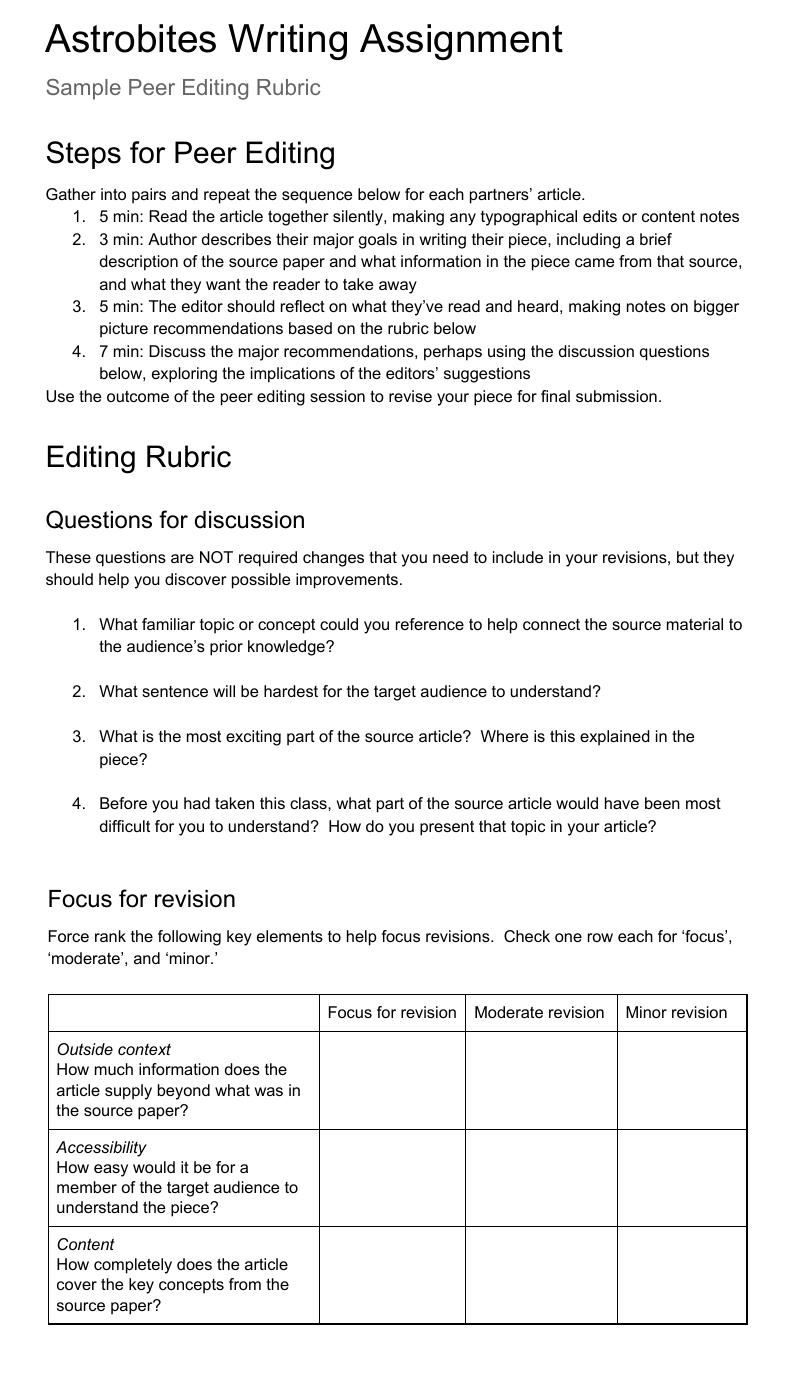}
\caption{Sample handout and rubric for peer editing for use with Method~3.}
\label{FigPeerEditing}
\end{figure}

\begin{itemize}
\item A project description sheet to explain the purpose and format of the writing assignment: \url{<https://goo.gl/qY3qt5>}

\item An article template for students to use as a starting point for their own written responses: \url{<https://goo.gl/uuHYOg>}

\item A peer editing rubric, illustrated in Figure~\ref{FigPeerEditing}, for students to use in workshopping their pieces: \url{<https://goo.gl/KOSIym>}
\end{itemize}

\begin{acknowledgments}

We acknowledge all the graduate student members of Astrobites who have contributed to our website and the governance of our collaboration since 2010.  We thank the students and educators who participated in our workshop at the 229th American Astronomical Society meeting in Grapevine, TX, whose feedback was instrumental to revising these lesson plans.  We are grateful to two anonymous reviewers for their thoughtful comments on this manuscript.  Astrobites is generously supported by the American Astronomical Society.

\end{acknowledgments}

\end{document}